\documentclass[a4paper,12pt]{extarticle}
\usepackage{geometry}
\usepackage[english]{babel}
\usepackage{amsmath,amssymb,amsfonts,amsthm}
\usepackage[mathscr]{eucal}
\usepackage[all]{xy}
\usepackage{bigints}
\usepackage{hyperref}
\usepackage{setspace}
\usepackage{upgreek}
\usepackage{bm}
\usepackage[center]{caption}
\usepackage{hyperref}
\usepackage{cite}
\usepackage{graphicx}
\usepackage{times}
\usepackage{color}

\usepackage{indentfirst}

\allowdisplaybreaks
\title{Coulomb problem for classical spinning particles}
\author{D.S. Kaparulin\thanks{E-mail: \texttt{dsc@phys.tsu.ru}}\; and
N.A. Sinelnikov\thanks{E-mail: \texttt{nikitasineln@gmail.com}}\\[0.5em]
{\normalsize Physics Faculty, Tomsk State University, Tomsk 634050, Russia}
}
\date{}

\begin{document}
\maketitle
\begin{abstract}
We consider the motion of a weakly relativistic charged particle with an arbitrary spin in central potential $e/r$ in terms of classical mechanics. We show that the spin--orbital interaction causes the precession of the plane of orbit around the vector of total angular momentum. The angular velocity of precession depends on the distance of the particle from the centre. The effective potential for in-plane motion is central, with the corrections to Coulomb terms coming from spin--orbital interaction. The possible orbits of a quantum particle are determined by the Bohr--Sommerfeld quantization rule. We give examples of orbits corresponding to small quantum numbers, which were obtained by numerical integration of equations of motion. The energies of stationary states are determined by spin--orbital interaction.
\end{abstract}
Since the works of Frenkel~\cite{Frenkel}, Mathisson~\cite{Mathisson} and Papapetrou~\cite{Papapetrou}, the spinning particle concept is used for the quasi-classical description of dynamics of spin in the electromagnetic and/or gravitational field. For a review of early results, we cite book~\cite{Corben}, and recent articles~\cite{Fryd,Deriglazov} for a modern exposition of the problem. The current applications of spinning particle theory include accelerator physics~\cite{ap-spin1,ap-spin2}, celestial mechanics~\cite{Adler}, laser physics~\cite{Karlovets}, statistical physics~\cite{BKN-2022}, astrophysics and cosmology~\cite{TFC-1976,Semerak,BCL-2008,OST-2013,Han-2017,FKR-2017,APP-2019,TM-2019,LL-2023}. In comparison with the approaches of relativistic quantum mechanics and quantum field theories, the quasi-classical approach has several advantages. The equations of motion include the spin as the continuous parameter that can take any non-negative real number, providing a description of particles of all spins on an equal basis. In quantum theory, the particles of different spins are described by different sets of dynamical variables~\cite{LL}. It is also known that the quasi-classical theory admits consistent couplings with the general electromagnetic and field~\cite{Univ} (see also~\cite{Deriglazov}). This property is especially important for higher-spin particles, where obstructions for admissible backgrounds are known in field theory already at the classical level~\cite{o-1,o-2}. The final advantage is that the quasi-classical equations allow the inclusion of anomalous magnetic moment of particles as a phenomenological parameter. This allows us to consider some effects of quantum field theory already at the mechanical level, even though the corresponding relativistic wave equations do not describe them. For example, the quasi-classical theory describes the precession of particle spin caused by the presence of anomalous magnetic moment of electron~\cite{DP-2014}. Another effect of the same order is a part of the Lamb shift of hydrogen atom energy levels associated with the anomalous magnetic moment of the electron. As we will see below, this part can be reproduced by the quasi-classical theory even though, it is not predicted by the Dirac equation. All the above-mentioned makes the spinning particle theory a useful tool for studying the dynamics of spinning particles. 

In the current article, we focus on studying two general aspects of spinning particle dynamics: the phenomenon of zitterbewegung and the application of quasi-classical quantization rules to particle trajectories. By zitterbewegung phenomenon, we mean the fast oscillations of classical positions that are observed in many models of spinning particles~\cite{Univ,Bratek,Rempel}. The article~\cite{WS} tells us that the oscillations of the classical path are the general properties of dynamics of free particles, whose quantization corresponds to the irreducible representation of the Poincare group. The amplitude of oscillations is determined by the model parameters. In the theory with interaction~\cite{Univ}, the motion of the free particle follows along the rectilinear path, while the inclusion of an external field causes another type of trajectory fluctuation. The effect was first observed in the integration of equations of motion in the uniform magnetic field in article~\cite{DP-2014} under the name magnetic zitterbewegung. The effect is physical because the particle position is observable in the model with coupling, and it is connected with the influence of particle spin on the translational motion of a particle. For a particle travelling with velocity $\beta c$ (with $c$ being the speed of light), the amplitude of the magnetic zitterbewegung is estimated as $\beta \lambda_C$, with $\lambda_C$ being the Compton wavelength. The effect is relativistic, as its magnitude tends to zero in the limit $c\to\infty$. The phenomenon of electric zitterbewegung has not been observed to date to our knowledge, though somewhat similar has been observed in~\cite{KR-2022} in a uniform electric field in $1+2$ dimensions. The other aspect concerns the determinations of quasi-classical stationary states of the particle by means of the Bohr--Sommerfeld quantization rules. Even though the model is well-tested for a scalar particle, we do not know its applications to the dynamics of particles with spin. This makes it interesting to test quasi-classical quantization rules in spinning particle theory. To study both interesting problems, we consider the motion of a spinning particle in the Coulomb field. For reasons of simplicity, we restrict ourselves to a weakly relativistic approximation.

The history of the study of the Coulomb problem for an electron in the hydrogen atom began before the discovery of spin. The quasi-classical trajectories of a point charged particle in the Coulomb field, were first considered by Bohr in~\cite{Bohr}. Here, the quantization of energy and angular momentum was proposed. The problem was reconsidered with an account of relativistic effects and elliptical motion by Sommerfeld~\cite{Sommerfeld}. Both authors considered a spinless particle, and their predictions agreed with the experiments. The quantum solution for the Coulomb problem for a non-relativistic spinless particle was found by Schrödinger in article~\cite{Schroedinger}. The predicted energy levels were the same as in the Bohr model. The Coulomb problem for a relativistic field of spin $1/2$ was solved by Dirac~\cite{Dirac}. Due to accidental coincidence, the energy levels of the model were given by the relativistic Sommerfeld formula, even though the systems are different (a spinning and a spinless particle). There is no contradiction between the results because the predictions of quasi-classical theory should be exact only in the limit of large quantum numbers. The relativistic wave equation for an arbitrary spin quantum particle in the Coulomb field was solved in~\cite{FN}. In the current study, we address three following questions: (i) to derive the equations of motion for a weakly relativistic particle; (ii) to integrate the equations of motions and find classical orbits; (iii) to apply the Bohr--Sommerfeld quantization rule to determine the quasi-classical orbits for the particle. We show the following results. First, we observe that spin causes the precession of the orbital plane of the particle around the vector of total angular momentum. This makes ``smashing'' of the particle trajectory, which can be considered as electrical zitterbewegung. Second, we find that the spin--orbit coupling changes the shape of the trajectory, making the orbit non-Keplerian. The frequency of the apse line precession is found to be a rational multiple of the orbital plane precession in the non-relativistic limit for the quasi-classical states satisfying Bohr--Sommerfeld quantization rules. This implies a special form of quasi-classical particle orbits resembling the Lissajous figures known in oscillation theory. Finally, we derive the energies of particle quasi-classical orbits. In the limit of large quantum numbers, the result agrees with the result of~\cite{FN} with the $o(\alpha^5)$ precision ($\alpha$ stands for the fine structure constant) for a particle of arbitrary spin. Accounting for the anomalous magnetic moment of the electron and its spin $s=1/2$, we find that the quasi-classical theory reproduces the part of Lamb shift associated with anomalous magnetic moment with the precision $\alpha^5$. As this part of the Lamb shift dominates for states with large orbital quantum numbers, the quasi-classical theory reproduces the energy of stationary states with the precision  $o(\alpha^5)$ for sufficiently large quantum numbers, better than the relativistic quantum mechanics can~do. 

The article is organized as follows. In the next section, we describe the classical dynamics of the model. We start with the presentation of the model of a weakly relativistic spinning particle travelling in the external electromagnetic field proposed in~\cite{Der-2016}. The equations of motion of the particle are derived in the first-order form and integrated in the analytical form. In Section \ref{sec3}, we apply the Bohr--Sommerfeld quantization rules to determine the stationary orbits of the particle in the Coulomb field. We conclude that the quasi-classical states are determined by four quantum numbers: the radial quantum number $n_r$, the total angular momentum quantum number $j$, the orbital quantum number $\ell$, and the magnetic quantum number $m$. The solution for the particle energy levels agrees with the formula of Fushich and Nikitin~\cite{FN} in the limit of large quantum numbers. At the end of the section, we present the results of the numerical simulations of quasi-classical orbits for selected quantum numbers. The conclusion summarizes the results.

\section{Classical Dynamics}
We consider the model of a point weakly relativistic massive charged spinning particle with mass $m$ and spin $s$ proposed in~\cite{Der-2016}. The particle state is determined by the particle position $\mathbf{x}$, particle linear momentum $\mathbf{p}$, and spin vector $\mathbf{s}$. The normalizing condition for spin vector reads $(\mathbf{s},\mathbf{s})=\hbar^2s(s+1)$. The round brackets determine the scalar product with respect to the Euclidean metric. The Poisson brackets on the phase space have the following form:
\begin{equation}
\begin{gathered}
    \{x^i,p^j\}=\delta^{ij}\,,\qquad \{s^i,s^j\}=\epsilon^{ijk}s^k\,;\\
    \{x^i,x^j\}=\{p^i,p^j\}=\{x^i,s^j\}=\{p^i,s^j\}=0\,.
    \end{gathered}
\end{equation}
Here, $x^i$, $p^i$, $s^i$, $i=1,2,3$ denote the components of the spatial vectors $\boldsymbol{x}$, $\boldsymbol{p}$, $\boldsymbol{s}$, and $\epsilon$ stands for the $3d$ Levi--Civita symbol. We use the convention $\epsilon^{123}=1$ throughout the paper. All the spatial indices $i,j,k,\ldots$ are raised and lowered by the Euclidean metric. The following Hamiltonian describes the motion of the charged spinning particle in the field of unit electric charge located at the origin, 
\begin{equation}\label{Ham0}
    H=\frac{1}{2m}\mathbf{p}^2-\frac{1}{8m^3c^2}\mathbf{p}^4-\frac{e^2}{r}+\frac{(g-1)e^2}{2m^2c^2r^3}(\boldsymbol{\mathbf{s}},\boldsymbol{\mathbf{L}})\,.
\end{equation}
Here, $m$ is the particle mass, $s$ is the particle spin, the orbital angular momentum is $\mathbf{L}=[\mathbf{x},\mathbf{p}]$. The constant $e$ is an elementary electric charge, and $c$ is the speed of light. The quantity $r$ denotes the distance from the origin.  

In the current article, we use a special unit system where the distances are measured in the units of Bohr radius, the momentum in the units of momentum on the Bohr orbit, energies in the particle rest energy, and spin in the Planck units. The new dynamical variables $\widetilde{\mathbf{x}}$, $\widetilde{\mathbf{p}}$, $\widetilde{\mathbf{s}}$ are introduced by the rule
\begin{equation}
    \mathbf{x}=r_B\widetilde{\mathbf{x}}=\frac{\hbar^2}{me^2}\widetilde{\mathbf{x}}\,,\qquad p=mv_B\widetilde{\mathbf{p}}=\frac{me^2}{\hbar}\widetilde{\mathbf{p}}\,,\qquad \mathbf{s}=\hbar\widetilde{\mathbf{s}}\,.
\end{equation}
($r_B$ and $v_b$ denote the Bohr radius and the speed on the first Bohr orbit). In the subsequent computations, we exclusively use the variables  $\widetilde{\mathbf{x}}$, $\widetilde{\mathbf{p}}$, and $\widetilde{\mathbf{s}}$, with the tilde omitted to avoid cumbersome notation. We rewrite the Hamiltonian (\ref{Ham0}) in the following form: 
\begin{equation}\label{Ham}
    H=\frac{\alpha^2}{2}\boldsymbol{\mathbf{p}}^2-\frac{\alpha^4}{8}\boldsymbol{\mathbf{p}}^4-\frac{\alpha^2}{r}+\frac{(g-1)}{2}\frac{\alpha^4}{r^3}(\boldsymbol{\mathbf{s}},\boldsymbol{\mathbf{L}})\,.
\end{equation}
Here, $\alpha=e^2/\hbar c$ is the fine structure constant. The Hamiltonian (\ref{Ham}) can be considered as the power series in the small parameter $\alpha$. The leading contribution in $\alpha$ corresponds to the non-relativistic Hamiltonian, and the quartic contribution accounts for the first relativistic correction. The exact expression for the Hamiltonian (\ref{Ham0}) is given by the infinite series in the inverse powers of the speed of light $1/c$. This corresponds to the infinite power series in $\alpha$ in (\ref{Ham}). The next-order correction will be proportional to $\alpha^6$. In the current article, we restrict ourselves with the precision $o(\alpha^4)$, with all the higher-order corrections being systematically ignored. This assumption is not a serious limitation because the quasi-classical theory is expected to be correct for energy levels with large quantum numbers. The relativistic corrections for these energy levels are expected to be small.

The Hamiltonian equations of motion for the dynamical variables $\mathbf{x}$, $\mathbf{p}$, $\mathbf{s}$ read
\begin{equation}\label{EoM}\left\{\begin{array}{l}\displaystyle
    \frac{d\mathbf{x}}{dt}=\bigg(1-\frac{\alpha^2}{2}\mathbf{p}^2\bigg)\mathbf{p}+\frac{(g-1)}{2}\frac{\alpha^2}{r^3}[\mathbf{s},\mathbf{x}]\,;\\[7mm]
    \displaystyle
    \frac{d\mathbf{p}}{dt}=-\bigg(\frac{1}{r^3}-\frac{3(g-1)}{2}\frac{\alpha^2}{r^5}(\mathbf{s},\mathbf{L})\bigg)\mathbf{x} +\frac{(g-1)}{2}\frac{\alpha^2}{r^3}[\mathbf{s},\mathbf{p}];\\[7mm]
    \displaystyle
    \frac{d\mathbf{s}}{dt}=\frac{(g-1)}{2}\frac{\alpha^2}{r^3}[\mathbf{L},\mathbf{s}]\,.
\end{array}\right.\end{equation}
Integration of these equations gives the particle trajectory with the initial position $\mathbf{x}_0$, initial momentum $\mathbf{p}_0$, and initial value of spin $\mathbf{s}_0$. The system has four obvious integrals of motion, the vector of total angular momentum $\boldsymbol{J}$, and one scalar quantity $\mathbf{L}^2$. As the spin vector is normalized, the latter also implies that the scalar product $(\mathbf{s},\mathbf{L})$ is also a constant on each classical trajectory. The physical meaning of the observation is that the vector of spin $\mathbf{s}$ and the vector of angular momentum $\mathbf{L}$ have precession around the vector of total angular momentum $\mathbf{J}$. This is confirmed by the fact that the equations of motion for $\mathbf{s}$, $\mathbf{L}$ can be represented in the form:
\begin{equation}
\label{LS-der}
    \frac{d\mathbf{L}}{dt}=[\mathbf{\Omega}_j,\mathbf{L}]\,,\qquad    \frac{d\mathbf{s}}{dt}=[\mathbf{\Omega}_j,\mathbf{s}]
    \,, \qquad \mathbf{\Omega}_{j}=\frac{(g-1)}{2}\frac{\alpha^2}{r^3}\mathbf{J}\,.
\end{equation}
\indent The angle $\gamma$ between vectors $\mathbf{J}$ and $\mathbf{L}$ is determined by the rule
\begin{equation}
\label{angle_gamma}
    \gamma=\arccos\bigg(\frac{\mathbf{J}^2+\mathbf{L}^2-\mathbf{s}^2}{2\sqrt{\mathbf{J}^2\mathbf{L}^2}}\bigg)\,,
\end{equation}
and it is a constant of motion. 

To find an exact solution to the equations of motion, it is convenient to introduce the non-inertial system (\ref{EoM}) rotating around the spin vector. Following~\cite{LLm}, we denote the time derivative in the rotating coordinate system by $d'/dt$. The derivative of an arbitrary vector, being a function on the phase space of variables $\mathbf{x}$, $\mathbf{p}$, $\mathbf{s}$ (and measured in non-inertial frame) reads
\begin{equation}
\frac{d}{dt}=\frac{d'}{dt}+
\bigg[\mathbf{\Omega}_s, \cdot\bigg]\,,\qquad \mathbf{\Omega}_s=\frac{(g-1)}{2}\frac{\alpha^2}{r^3}\mathbf{s}\,.
\end{equation} 
\indent In the rotating coordinate system, Equation (\ref{EoM}) take the following form:
\begin{equation}\label{EoM-rot}
\left\{\begin{array}{c}\displaystyle
\frac{d'\mathbf{x}}{dt}=\bigg(1-\frac{\alpha^2}{2}\mathbf{p}^2\bigg)\mathbf{p}\,,\qquad \frac{d'\mathbf{p}}{dt}=-\bigg(\frac{1}{r^3}-\frac{3(g-1)}{2}\frac{\alpha^2}{r^5}(\mathbf{s},\mathbf{L})\bigg)\mathbf{x}\,;\\[7 mm]
    \displaystyle
    \frac{d'\mathbf{s}}{dt}=\frac{(g-1)}{2}\frac{\alpha^2}{r^3}[\mathbf{L},\mathbf{s}]\,.
\end{array}\right.\end{equation}
\indent The first and second equations describe the dynamics of the translational degrees of freedom. They correspond to the motion in the central field with the effective potential
\begin{equation}\label{U-eff}
    U_{\text{eff}}=-\frac{\alpha^2}{r}+\frac{(g-1)}{2}\frac{\alpha^4}{r^3}(\mathbf{s},\mathbf{L})\,.
\end{equation}
\indent In the last expression, the scalar product $(\mathbf{s},\mathbf{L})$ is considered as the parameter, being independent of the radial variable. The radial potential (\ref{U-eff}) is given by the sum of the Coulomb term and spin--orbital contribution, being a small correction. The translational motion in the rotating coordinate system is planar, as the orbital angular momentum is conserved in the non-inertial frame. The last equation in the system (\ref{EoM-rot}) describes the precession of spin, and it has the original form (\ref{EoM}). Equations (\ref{EoM-rot}) are important because they allow us to decouple the dynamics of translational and spin degrees of freedom. 
After this, both degrees of freedom can be considered independently.   

Let us consider translational motion. Without loss of generality, we assume that the vector of orbital angular momentum $L$ is directed along the third coordinate axis of the non-inertial frame, so the motion occurs in the orthogonal to $\mathbf{L}$ plane. We describe the particle position in the plane by polar coordinates $r,\varphi$. The system has two obvious quantities preserved by the planar motion: the norm of orbital angular momentum $L$, and the energy of translational motion. The first relation determines the angular velocity of motion around the origin in the second-Kepler-law style, 
\begin{equation}\label{L-df}
    \frac{d'\varphi}{dt}=\frac{L}{r^2}\,.
\end{equation}
\indent The law of conservation of energy determines the velocity of radial motion. Taking into account that
\begin{equation}
    \mathbf{p}^2=\bigg(\frac{d'r}{dt}\bigg)^2+\frac{\mathbf{L}^2}{r^2}\,,
\end{equation}
with the precision $o(\alpha^4)$, we obtain
\begin{equation}\label{E-dr}
    \frac12\bigg(\frac{d'r}{dt}\bigg)^2=\frac{1}{\alpha^2}\Bigg[\bigg(E+\frac{\alpha^2}{r}\bigg)+\frac{1}{2}\bigg(E+\frac{\alpha^2}{r}\bigg)^2-\frac{\alpha^2L^2}{2r^2}-\frac{(g-1)}{2}\frac{\alpha^4}{r^3}(\mathbf{s},\mathbf{L})\Bigg]\,.
\end{equation}
\indent Combining (\ref{L-df}) and (\ref{E-dr}), we obtain two ODEs with respect to $\varphi(r)$ and $t(r)$. Their solution reads
\begin{equation}\label{EoMpolar}
t-t_0=\pm \bigintsss_{r_0}^r\frac{\alpha dr}{\displaystyle
\sqrt{2\bigg(E+\frac{\alpha^2}{r}\bigg)+\bigg(E+\frac{\alpha^2}{r}\bigg)^2-\frac{\alpha^2L^2}{r^2}-\frac{(g-1)\alpha^4}{r^3}(\mathbf{s},\mathbf{L})}}\,;
\end{equation}
\begin{equation}\label{EoMpolar-1}
\varphi-\varphi_0=\pm \bigintsss_{r_0}^r\frac{\alpha L r^{-2}dr}{\displaystyle
\sqrt{2\bigg(E+\frac{\alpha^2}{r}\bigg)+\bigg(E+\frac{\alpha^2}{r}\bigg)^2-\frac{\alpha^2L^2}{r^2}-\frac{(g-1)\alpha^4}{r^3}(\mathbf{s},\mathbf{L})}}\,;
\end{equation}
\indent Both integrals admit an exact solution in terms of the Weierstrass elliptic functions. The explicit expressions for the integrals are long and not very informative, and thus do not present them here. Here, we restrict ourselves to the periods of integrals, that have the sense of the orbital period of the particle, and the angle of rotation between the two turning points of one type of radial motion. Both quantities are easily computed using the techniques presented in~\cite{LLm},
\begin{equation}\label{per}
    T=\frac{\pi \alpha^3}{\sqrt{-2E^3}}\bigg(1+\frac{1}{4}E\bigg)\,,\qquad \Phi=2\pi+\frac{\pi\alpha^2}{L^2}\bigg(1-3(g-1)\frac{(\mathbf{s},\mathbf{L})}{L^2}\bigg)\,.
\end{equation}
{Here, the} quantity $\Delta \Phi_a=\Phi-2\pi$ determines the rotation of the apse line per one radial oscillation (between two pericentres or two apocentres). 

We finalize the section by returning to the inertial frame, where the particle orbit is involved in two precessions: the rotation of the orbital plane (\ref{LS-der}) and rotation of the apse \mbox{line (\ref{per}).} To compare the (average) angular velocities of precession, we compute the rotation angle for orbital motion per one radial oscillation,
\begin{equation}
   \Delta\Phi_{j}=\int^{T}_{0}\Omega_{j}(r)dt=\frac{\pi(g-1)\alpha^2J}{L^3}\,. 
\end{equation}
\indent The ratio of these two angles reads
\begin{equation}
\label{prec_ratio}
    \frac{\Delta \Phi_{a}}{\Delta \Phi_{j}}=\frac{L}{J}\bigg(\frac{1}{g-1}-\frac{3(\mathbf{s},\mathbf{L})}{L^2}\bigg)\,.
\end{equation}
\indent Even though this ratio can be an arbitrary real number it is a rational quantity for quasi-classical orbits for a particle without anomalous magnetic moment. Indeed, the quasi-classical quantization rules imply that $L$ and $J$ are integer (integer and half-integer) numbers (see Section \ref{sec3}), while the scalar product $(\mathbf{s},\mathbf{L})$ is rational because of Formula (\ref{angle_gamma}). Finally, for a particle without anomalous angular momentum, $g=2$ and $g-1=1$. To our knowledge, {the fact that} the precession velocities of the apse line and orbital plane are rational multiples {has} has been observed here for the first time. 
It is important to note that the quantity (\ref{prec_ratio}) is rational only in the non-relativistic limit. The relativistic effects make this ratio irrational even for quasi-classical orbits.

\section{Quasi-Classical Orbits}
\label{sec3}

The Coulomb problem for a relativistic quantum particle is known to be an exactly solvable model, with the solution provided in~\cite{FN}. In this section, we reproduce the solution for the energies of stationary states by applying the Bohr--Sommerfeld quantization rule for the weakly relativistic particle and compare the result with the previously known exact solution.

We start with the consideration of the rotational motion. The symmetry of the \mbox{model (\ref{EoM})} has three following integrals of motion in involution: $\mathbf{J}^2$, $\mathbf{L}^2$ and $J_z$. The quantization of orbital angular momentum follows from the so-called ``spatial quantization'' rules proposed in~\cite{Sommerfeld}. We use the quantization condition for $\mathbf{L}^2$ because $\mathbf{L}$ is not conserved in the model. The quantization of total angular momentum follows from the fact that the spin projection on a fixed axis takes (half-)integer numbers, while the classical vector $\mathbf{J}$ is directed along some vector. In its turn, the quantization of spin in quasi-classical theory is predicted by the co-orbit method~\cite{Kirillov,Kostant,Souriau}. Both mentioned conditions imply that $J_z$ takes the (half-)integer numbers, while $\mathbf{J}^2$ and $\mathbf{L}^2$ are given by the squares of (half-)integer and integer numbers,
\begin{equation}
\label{stand_quant}
    \phantom{\frac12}\mathbf{J}^2=j^2\,,\qquad \mathbf{L}^2=\ell^2\,,\qquad J_z= m\,.\phantom{\frac12}
\end{equation}
(we recall that $\hbar=1$ in our unit system). The quantum numbers are the total angular momentum quantum number $j$, the orbital quantum number $\ell$, and the magnetic quantum number $m$. The quantum numbers are restricted by the condition that the vectors $\mathbf{J}$, $\mathbf{L}$ and $\mathbf{s}$ form a triangle with normalized $s$. Quantum numbers $j$, $\ell$, $m$ run over the set
\begin{equation}\label{jl-quant}
    \phantom{\frac12}\ell=1,2,\ldots,\qquad m=-j,\ldots,j\,,\qquad j=\ell-m_{s\ell},\ldots,\ell+s \,.\phantom{\frac12}
\end{equation}
where $m_{s\ell}=\text{min}\{s,\ell\}$. The quantization rules for the vectors $\mathbf{J}^2$, $\mathbf{L}^2$, and $J_z$ are universal for any particle model with the spherical symmetry of the potential and are not connected with the specifics of the model. 

The non-trivial condition comes from the discretization of radial motion. The quantization rule for radial variable reads
\begin{equation}\label{nr}
    \int p_r dr=2\pi n_r\,,\qquad n_r=0,1,\ldots\,.
\end{equation}
\indent Expressing the radial momentum from the Hamiltonian (\ref{Ham}), we arrive at the following condition:
\begin{equation}\label{nr1}
   \bigointsss \alpha^{-1}\sqrt{2\bigg(E+\frac{\alpha^2}{r}\bigg)+\bigg(E+\frac{\alpha^2}{r}\bigg)^2-\frac{\alpha^2L^2}{r^2}-\frac{(g-1)\alpha^4}{r^3}(\mathbf{s},\mathbf{L})}\; dr =2\pi n_r\,,
\end{equation}
where $E$ is the energy of the particle, being negative for finite motion. Calculating the integral (\ref{nr1}) with the precision $o(\alpha^4)$ and expressing energy, we obtain 
\begin{equation}\label{energy_njl}
\displaystyle
       E_{njl}=-\frac{\alpha^2}{2n^2}-\frac{\alpha^4}{2n^4}\left(\frac{n}{\ell}-\frac{3}{4}-(g-1)\frac{n(j^2-\ell^2-s(s+1))}{2\ell^3}\right)\,, 
\end{equation}
with $n=n_r + \ell$ being the principal quantum number. The first term is the Bohr energy, and the second is the first relativistic correction to energy. When $s=0$, so $j=\ell$, this formula coincides with the Sommerfeld one (and the quantum levels of the Dirac particle). For a particle with the general value of spin, energy depends on three quantum numbers, except $m$. This means that the account of spin decreases the degeneracy of energy levels in the Coulomb problem. This fact has also been observed in~\cite{FN}. 

Comparing the values of energy levels with the energies of the quantum state derived from the solution of relativistic wave Equation~\cite{FN}, we note that the coincidence of the results (even in the order $o(\alpha^4)$) will be a too strong requirement. Indeed, the classical relativistic Sommerfeld formula~\cite{Sommerfeld} reproduces the energy levels of the Dirac particle~\cite{Dirac} with spin $s=1/2$. The Sommerfeld formula agrees with the energies of the quantum scalar particle only in the limit of large quantum numbers $\ell,n\to\infty$. Thus, the requirement of large quantum numbers seems to be necessary for comparing quantum and quasi-classical results. One more condition 
includes the spin value. We account for the spin--orbit term by the perturbation theory, even though this contribution is proportional to spin, being an unlimited number. For extremely high spins, the perturbation theory no longer applies. To avoid contradiction, we need to assume that spin $s$ is small in comparison with other quantum numbers, i.e., $n,\ell\gg s\gg1$. Finally, the work~\cite{FN} considers the special value of the magnetic moment of the particle, while in our case the $g$-factor is an arbitrary number. 

Now, we can compare Formula (\ref{energy_njl}) with the predictions of quantum theory. The work~\cite{FN} gives the following expression for the energy levels with the precision $o(\alpha^4)$:
\begin{equation}\label{energy_FN}
    E_{njl}=-\frac{\alpha^2}{2n^2}-\frac{\alpha^4}{2n^4}\bigg(\frac{2n}{2\ell+1}-\frac{3}{4}-\frac{1}{4s^2}\frac{n}{2\ell+1}\bigg(\frac{a^{sj}_{j-\ell+s}}{\ell+1}-\frac{a^{sj}_{j-\ell+s+1}}{\ell}\bigg)\bigg)\, , 
\end{equation}
where
\begin{equation}
    a^{sj}_{\mu}=\frac{\mu(2j-\mu+1)(2s-\mu+1)(2j+2s-\mu+2)}{(2j+2s-2\mu+1)(2j+2s-2\mu+3)}\,, 
\end{equation}
with $\mu$ being an arbitrary half-integer number. The leading term, being proportional to $\alpha^2$ is the same in both formulas. This is the energies of Bohr orbits in non-relativistic approximation. As for $\alpha^4$ correction, the first and second contributions account for relativistic effects. They become equal in Formulas (\ref{energy_njl}) and (\ref{energy_FN}) for large quantum numbers. The comparison for the third spin-dependent term requires an accurate computation of the coefficients $a^{sj}_{\mu}$ in the large quantum number limit (in the above-described sense). The identification of two formulas follows from the fact that 
\begin{equation}
\frac{1}{2\ell+1}\bigg(\frac{a^{sj}_{j-\ell+s}}{\ell+1}-\frac{a^{sj}_{j-\ell+s+1}}{\ell}\bigg)\approx\frac{(j^2-\ell^2-s(s+1))}{2\ell^3}\approx \frac{j-\ell}{l^2}\,.
\end{equation}
\indent Both formulas give the same result for a spin for a special value of gyromagnetic ratio $g=1+1/4s^2$. For a spin $s=1/2$ particle, this corresponds to a particle without anomalous magnetic moment, $g=2$. To make further progress, we restrict ourselves with spin $s=1/2$ particle. Using the estimate for the anomalous magnetic moment of electron $(g-2)/2\approx \alpha/2\pi$~\cite{PS}, we find the energy (\ref{energy_njl}) in the following form:
\begin{equation}\label{energy_njl_2}
\displaystyle
       E_{njl}=-\frac{\alpha^2}{2n^2}-\frac{\alpha^4}{2n^4}\left(\frac{n}{\ell}-\frac{3}{4}-\frac{n(j^2-\ell^2-3/4)}{2\ell^3}\right)+\frac{\alpha^5}{4\pi n^3}\frac{(j^2-\ell^2-3/4)}{\ell^3}\,. 
\end{equation}
\indent In the limit of large quantum numbers $n,j,\ell\gg1$, we obtain:
\begin{equation}\label{energy_njl_3}
\displaystyle
       E_{njl}=-\frac{\alpha^2}{2n^2}-\frac{\alpha^4}{2n^4}\left(\frac{n}{j}-\frac{3}{4}\right)+\frac{\alpha^5(j-\ell)}{2 \pi n^3\ell^2}\,. 
\end{equation}
Here, the $\alpha^4$ term corresponds to the Dirac correction to the energy of stationary states, accounting for both spin and relativistic effects. The $\alpha^5$ term reproduces part of the Lamb shift associated with the anomalous magnetic moment of an electron. The expression is exact in the limit of large orbital quantum numbers~\cite{BS}. The complete expression for the Lamb shift includes the part connected with the polarization of the vacuum near the origin. This effect of the quantum field theory is not taken into account by the quasi-classical model. However, the Bethe logarithms are ``extremely small'' for large values of $\ell$~\cite{BS}, so the anomalous magnetic moment determines the leading part of the Lamb shift for highly excited levels. Summarizing the above, we see that the quasi-classical theory (with an account of the anomalous magnetic moment) gives the correct formula for the energies of highly excited stationary states of a hydrogen atom with the precision $o(\alpha^5)$, better than the relativistic quantum mechanics can do. 

We finalize the section by discussing the properties of quasi-classical trajectories for various quantum numbers. The states with $n_r=0$ (or equivalently, $n=\ell$) correspond to circular orbits. The radius of the orbit is determined by the minimum of the effective potential of radial motion entering the right-hand side of Equation (\ref{E-dr}). Finding the extremum of the function, we find the radii of circular orbits with the precision $o(\alpha^2)$, 
\begin{equation}
    r_{\ell\ell j}=\ell^2-\frac{\alpha^2}{2}-\frac{3\alpha^2(g-1)}{4}\frac{(j^2-\ell^2-s(s+1))}{\ell^2}\,.
\end{equation}
\indent The first contribution is the classical radius of the Bohr orbit, the second accounts for relativistic correction, and the third contribution accounts for a spin. The expression depends on both the quantum numbers $j$ and $l$, so we have no {degeneracy} by the orbital quantum number having a place in relativistic quantum mechanics. The orbital plane precession ``smashes'' the circular orbits into ring-like spherical segments with width \mbox{2$\gamma$ (\ref{angle_gamma}).} In the large quantum number limit, we have an estimate for the linear width of the segment, 
\begin{equation}
    h_{\ell j}=2r_{\ell\ell j}\gamma_{\ell\ell j}\approx 2\ell \sqrt{s(s+1)-(j-\ell)^2}.
\end{equation}
\indent The state with the minimal possible projection of spin corresponds to the widest spherical segment. The smashing of particle circular orbit can be considered as the spherical form of electrical zitterbewegung somewhat similar to magnetic zitterbewegung introduced in~\cite{DP-2014}.

\begin{figure}[h!]
  
  \includegraphics[width=1\linewidth]{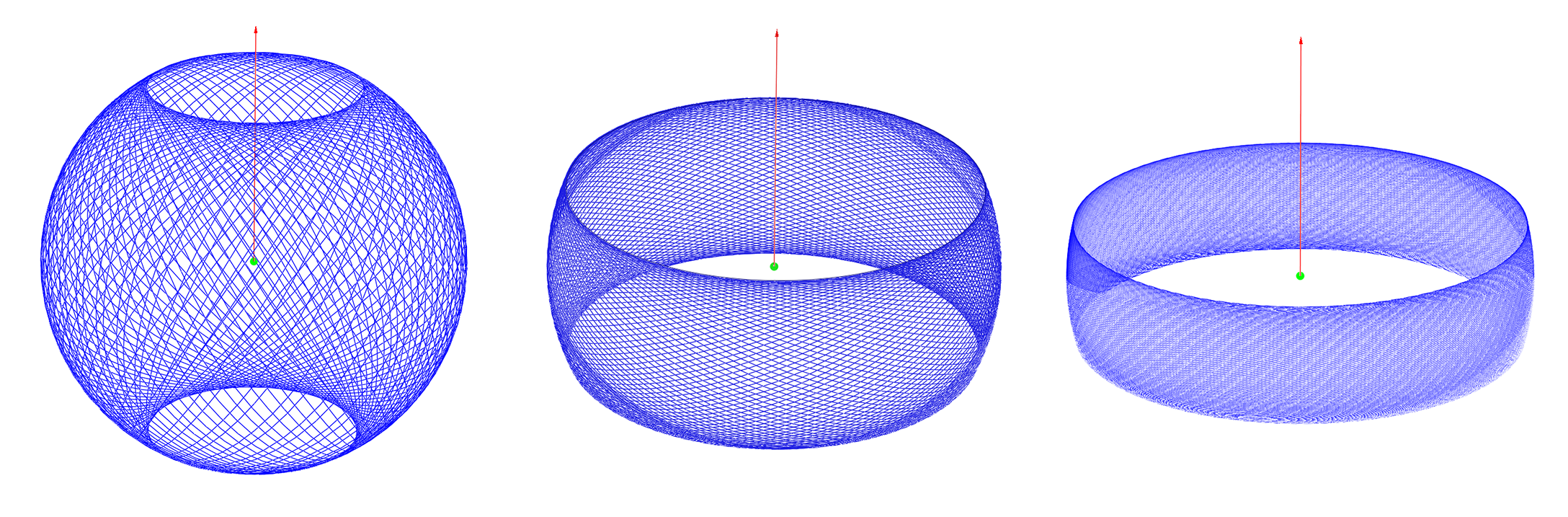}
  \centering
  \captionof{figure}{Orbits of an electron for zero radial quantum number. From left to right:  $\ell=1, \; j=1/2;$ $ \ell=2, \; j=3/2;~\ell=3, \; j=5/2.$}
  \label{fig1}
\end{figure}

\begin{figure}[h!]
\centering
  \includegraphics[width=0.5\linewidth]{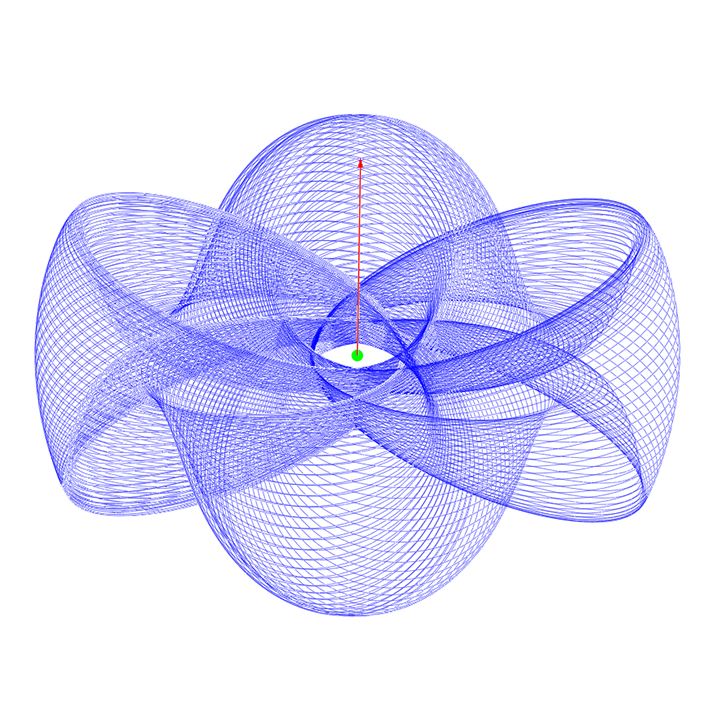}
  \captionof{figure}{An elliptical orbit for an electron with quantum numbers: $n_r = 1, \; \ell = 2, \; j=5/2.$}
  \label{fig5}
\end{figure}

To visualize orbits, the equations of motions (\ref{EoM}) were numerically integrated for selected quantum numbers to demonstrate features of quasi-classical trajectories of an electron with spin $1/2$. 
We have chosen small quantum numbers and increased the value of the $g$-factor $g=2000$ to increase the precession speed to get more illustrative images. For large quantum numbers, we have {ring-shaped spherical segments} whose height is much smaller than the radius. The green point is the position of the Coulomb attraction centre. The red arrow represents the vector of the total angular momentum. Figure \ref{fig1} represents the circular orbits with zero radial quantum number $n_r$. For a non-zero radial quantum number, there is a radial motion, the pericentre and apocentre distances being determined by the turning points of the potential (\ref{E-dr}). There are two factors affecting the orbit: the precession of the orbit plane and the precession of the pericentre of the orbit. \mbox{Formula (\ref{prec_ratio})} tells us that the ratio of the frequencies of precession is a rational number for each quasi-classical state whenever the $g$-factor is a rational quantity. Figure \ref{fig5} shows us that the classical path forms the Lissajous-like figures due to this effect. The small quantum numbers are chosen for illustrative purposes. At a very long integration time, the effect disappears due to the approximate character of the ratio of angular velocities of precession. Therefore, at the large time scales the classical path of the spinning particle draws a figure consisting of spherical segments of angular width $2\gamma$ (\ref{angle_gamma}) and radius changing between peri- and apocentre distances. In the limit of extremely large quantum numbers, we almost have flat trajectories resembling the usual Bohr orbits.

\vspace{-6pt}

\newpage

\section{Conclusions}

In the current article, we have considered the Coulomb problem for a weakly relativistic particle of arbitrary spin from the quasi-classical point of view. Using the results of~\cite{Der-2016}, we derived the equations of motion of the particle in the field of a point charge and integrated them. It has been shown that the presence of spin affects the particle motion in two ways. It causes the precession of the orbital plane of the particle at first, and it generates the precession of the apse line, making the in-plane motion non-Keplerian. The ratio of precession angular velocities was found to be rational for each quasi-classical state. Using the Bohr--Sommerfeld quantization rule, we found the energies of stationary states in the quasi-classical formalism with the precision {$o(\alpha^5)$}. Our formula agrees with the quantum mechanical result of Fushich and Nikitin~\cite{FN} in the limit of large quantum numbers for a particle without anomalous magnetic moment. {For an electron with spin $s=1/2$, we calculated a more precise estimate. Using the expression for the anomalous magnetic moment $g-2=\frac{\alpha}{\pi}+o(\alpha)$, the $\alpha^5$ correction to the energies of stationary states was found. We identified part of the Lamb shift associated with electron anomalous magnetic moment.
	This part dominates for high quantum numbers, so the quasi-classical theory gives the correct estimate for the Lamb shift for states with large orbital quantum numbers.} Finally, we made the visualization of quasi-classical orbits for small quantum numbers. The orbits with zero radial quantum numbers lie on spherical segments whose width depends on the total angular momentum and orbital quantum number. This shape of circular obits better resembles the spherically symmetric quantum states that the planar Bohr orbits. The non-planar motion of an electron can be considered a special form of zitterbewegung. The results of the article demonstrate once again that a spinning particle concept is a useful tool for the study of various phenomena involving quantum particles, including the motion in the external field.
\\\\
\textbf{Acknowledgments.} The authors thank the hospitality of organizers of the "AYSS-2022" conference of young scientists and specialists at JINR (Dubna, Russia) where preliminary results of the study have been presented by N.A. Sinelnikov. The authors thank A.A. Sharapov and N. Makhaldiani for valuable discussions of this work. We express our special gratitude to Yu.V. Brezhnev who participated in the initial stages of the project. The work was supported by the RSF project 19-71-10051-P.

{}

\begin{thebibliography}{}

\bibitem{Frenkel}  Frenkel, J. Die Elektrodynamik des rotierenden Elektrons. {\em Z. Phys.} {\bf 1926}, {\em 37}, 243--262. 

\bibitem{Mathisson} Mathisson, M. Neue mechanik materieller systeme. {\em Acta Phys. Pol.} {\bf 1937}, {\em 6}, 163--200.


\bibitem{Papapetrou} Papapetrou, A. Spinning test-particles in general relativity. {\em Proc. R. Soc. A} {\bf 1951}, {\em 209}, 248--258.


\bibitem{Corben}
Corben, H.C. {\em Classical and Quantum Theories of Spinning Particles}; Holden-Day: San Francisco, CA, USA, 1968; 279p.


\bibitem{Fryd} Frydryszak, A. Lagrangian models of the particles with spin: The first seventy years. In {\em From Field Theory to Quantum Groups}; World Scientific Publishing: Singapore, 1996; pp. 151--172.

\bibitem{Deriglazov} Deriglazov, A.A.; Ramírez, W.G. Recent progress on the description of relativistic spin: Vector model of spinning particle and rotating body with gravimagnetic moment in General Relativity. {\em Adv.  Math. Phys.} {\bf 2017}, {\em 2017}, 7397159. 

\bibitem{ap-spin1} Hoffstaetter, G.H.; Dumas, H.S.; Ellison, J.A. Adiabatic invariance of spin--orbit motion in accelerators. {\em Phys. Rev. ST Accel. Beams} {\bf 2006}, {\em 9}, 014001.

\bibitem{ap-spin2} Miller, J.P.; de Rafael, E.;  Roberts, B.L. Muon (g-2): Experiment and theory. {\em Rep. Prog. Phys.} {\bf 2007}, {\em 70}, 795.

\bibitem {Adler} Adler, R.J. The three-fold theoretical basis of the Gravity Probe B gyro precession calculation. {\em Class. Quant. Grav.} {\bf 2015}, {\em 32}, 224002.

\bibitem{Karlovets} Karlovets, D.V. Electron with orbital angular momentum in a strong laser wave. {\em Phys. Rev. A} {\bf 2012}, {\em 86}, 062102.

\bibitem{BKN-2022} Bubenchikov, M.A.; Kaparulin, D.S.; Nosyrev, O.D. Chiral effects in classical spinning gas. {\em J. Phys. A Math. Theor.} {\bf 2022}, {\em 55}, 395006. 

\bibitem{TFC-1976}  Tod, K.P.; de Felice, F.; Calvani, M. Spinning test particles in the field of a black hole. {\em Nuovo C. B} {\bf 1976}, {\em 34}, 365--379.

\bibitem{Semerak} Semerak, O. Spinning test particles in a Kerr field. {\em Mon. Not. R. Astron. Soc.} {\bf 1999}, {\em 308}, 863--875. 

\bibitem{BCL-2008}
Bastianelli, F.; Corradini, O.; Latini, E. Spinning particles and higher spin fields on (A)dS backgrounds. {\em JHEP} {\bf 2008}, {\em 2008}, 11. 

\bibitem{OST-2013} Obukhov, Y.N.; Silenko, A.J.; Teryaev, O.V. Spin in an arbitrary gravitational field. {\em Phys. Rev. D} {\bf 2013}, {\em 88}, 084014. 


\bibitem{Han-2017}  Han, W.-B.; Yang, S.-C. Exotic orbits due to spin–spin coupling around Kerr black holes. {\em Int. J. Mod. Phys. D} {\bf 2017}, {\em 27}, 1750179. 

\bibitem{FKR-2017}
Franciolini, G.; Kehagias, A.; Riotto, A. Imprints of Spinning Particles on Primordial Cosmological Perturbations. {\em JCAP} {\bf 2017}, \mbox{{\em 2}, 023.} 

\bibitem{APP-2019} Antoniou, I.; Papadopoulos, D.; Perivolaropoulos, L. Spinning particle orbits around a black hole in an expanding background.
{\em Class. Quantum Grav.} {\bf 2019}, {\em 36}, 085002.

\bibitem{TM-2019}
Toshmatov, B.; Malafarina, D. Spinning test particles in the $\gamma$ spacetime. {\em Phys. Rev. D} {\bf 2019}, {\em 100}, 104052. 

\bibitem{LL-2023}
Ladino, J.M.; del Valle, C.A.; Larranaga, E.
Motion of Spinning Particles around Black Holes. In {\em
A Guide to Black Holes}; Arun, K., Ed.; Nova Science Publishers: New York, NY, USA, 2022; pp. 79--96.


\bibitem{LL} Landau, L.D.; Lifshitz, E.M. {\em Quantum Mechanics. Non-Relativistic Theory,} 2nd ed.;  Pergamon Press: Oxford, UK, 1977; 632p.

\bibitem{Univ}
Lyakhovich, S.L.; Segal, A.Y.; Sharapov, A.A.  Universal model of a D = 4 spinning
particle. {\em Phys. Rev. D} {\bf 1996}, {\em 54}, 5223--5238. 

\bibitem{o-1}
Cortese, I.; Rahman, R.; Sivakumar, M. Consistent non-minimal couplings of massive higher
spin particles. {\em Nucl. Phys. B} {\bf 2014}, {\em 879}, 143--161. 

\bibitem{o-2}
Boulanger, N.; Deffayet, C.; Garcia-Saenz, S.; Traina, L. Consistent deformations of free massive
field theories in the Stueckelberg formulation. {\em JHEP} {\bf 2018}, {\em 1807}, 021. 

\bibitem{DP-2014} Deriglazov, A.A.;  Pupasov-Maksimov, A.M. Frenkel electron on an arbitrary electromagnetic background and magnetic Zitterbewegung. {\em Nucl. Phys. B} {\bf 2014}, {\em 885}, 1--24. 

\bibitem{Bratek} Bratek, L. Fundamental relativistic rotator: Hessian singularity and the issue of the minimal  interaction with electromagnetic field. {\em J. Phys. A Math.  Theor.} {\bf 2011}, {\em 44}, 195204. 

\bibitem{Rempel}
Rempel, T.;  Freidel, L. Bilocal model for the relativistic spinning particle. {\em Phys. Rev. D} {\bf 2017}, {\em 95}, 104014. 

\bibitem{WS}
Kaparulin, D.S.; Lyakhovich, S.L.World shetts of spininning particles. {\em Phys. Rev. D} {\bf2017}, {\em 96}, 105014. 



\bibitem{KR-2022} Kaparulin, D.S.; Retuntsev, I.A. On the world sheet of anyon in the external electromagnetic field.
{\em Nucl. Phys. B} {\bf 2022}, {\em 980}, 115836. 

\bibitem{Bohr} Bohr, N. On the Constitution of Atoms and Molecules. {\em Phil. Mag. Ser.} {\bf 1913}, {\em 6}, 1--24. 

\bibitem{Sommerfeld} Sommerfeld, A. \emph{Atombau und Spektrallinien};  F. Vieweg \& Sohn: Braunschweig, Germany, 1921.

\bibitem{Schroedinger} Schrödinger, E. Quantisierung als Eigenwertproblem. {\em Annalen Phys.} {\bf 1926}, {\em 386}, 109--139. 

\bibitem{Dirac}  Dirac, P.A.M. The quantum theory of the electron. {\em Proc. Roy. Soc. Lond. A} {\bf 1928}, {\em 117}, 610--624.

\bibitem{FN} Fushchich, W.I.; Nikitin, A.G. {\em Symmetries of Equations of Quantum Mechanics}; Allerton Press Inc.: New York, NY, USA, 1994; 480p.

\bibitem{Der-2016} Deriglazov, A.A.; Pupasov-Maximov, A.M. Relativistic corrections to the algebra of position variables and spin--orbital interaction. {\em Phys. Lett. B.} {\bf 2016}, {\em 761}, 207--212. 

\bibitem{LLm} Landau, L.D.; Lifshitz, E.M. {\em Mechanics}, 2nd ed.;  Pergamon Press: Oxford, UK, 1969; 176p.

\bibitem{Kirillov} Kirillov, A.A. {\em Elements of the Theory of Group Representations}; Springer: Berlin, Germany, 1976; 330p.

\bibitem{Kostant} Kostant, B. Quantization and unitary representations. In {\em Lectures in Modern Analysis and Applications III}; Lecture Notes in Mathematics; Springer: Berlin, Germany, 1970;  Volume 170,  pp. 87--208.

\bibitem{Souriau} Souriau, J.M. {\em Structure of Dynamical Systems: A Symplectic View of Physics}; Birkhauser: Basel, Switzerland, 2012; 426p.


\bibitem{PS}
Peskin, M.E.; Schroeder, D.V.  {\em An Introduction to Quantum Field Theory}; Addison-Wesley: Boston, MA, USA, 1995; 865p.

\bibitem{BS}
Bethe, H.A.; Salpeter, E.E. {\em Quantum Mechanics of One- and Two-Electron Atoms}; Springer: Berlin, Germany, 1957; 368p.

\end{thebibliography}
\end{document}